# 3D Channel Model in 3GPP

Bishwarup Mondal, Timothy A. Thomas, Eugene Visotsky, Frederick W. Vook, Amitava Ghosh, Nokia Networks, USA

Young-Han Nam, Yang Li, Charlie Zhang, Samsung Research America

Min Zhang, Qinglin Luo, Alcatel-Lucent, Alcatel-Lucent Shanghai Bell

Yuichi Kakishima, Koshiro Kitao, NTT DOCOMO, INC.

*Abstract* — Multi-antenna techniques capable of exploiting the elevation dimension are anticipated to be an important air-interface enhancement targeted to handle the expected growth in mobile traffic. In order to enable the development and evaluation of such multi-antenna techniques, the 3rd generation partnership project (3GPP) has recently developed a 3-dimensional (3D) channel model. The existing 2-dimensional (2D) channel models do not capture the elevation channel characteristics lending them insufficient for such studies. This article describes the main components of the newly developed 3D channel model and the motivations behind introducing them. One key aspect is the ability to model channels for users located on different floors of a building (at different heights). This is achieved by capturing a user height dependency in modelling some channel characteristics including pathloss, line-of-sight (LOS) probability, etc. In general this 3D channel model follows the framework of WINNERII/WINNER+ while also extending the applicability and the accuracy of the model by introducing some height and distance dependent elevation related parameters.

## I. INTRODUCTION

It is predicted that the traffic carried by wireless and mobile communication systems will increase around 1000-fold between 2010 and 2020. In addition there will be a proliferation of the number of connected devices and the communication systems will experience a coexistence of a diverse range of communication links ranging from very high data-rate mobile multimedia services to low data-rate machine-to-machine type communication [1]. In view of this, the International Telecommunications Union (ITU's) working party 5D is developing a recommendation on the framework and objectives of the future development of International Mobile Telecommunications (IMT) for 2020 and beyond.

Multi-antenna transmission techniques using massive antenna configurations (such as more than 8 transmit antenna ports) and exploiting 3D spatial dimensions (azimuth and elevation) of a multiple-input multiple-output (MIMO) channel are critical in improving the spectral efficiency and reliability of a radio-link. As a result of the significant interest shown in 3GPP in supporting advanced MIMO transmission techniques exploiting both azimuth and elevation dimensions of a wireless channel, a study was started in January 2013 targeting the incorporation of a 3D channel model into the 3GPP evaluation methodology [2]. In this article we describe the main elements of the finalized 3D channel model and also provide the motivations for the various decisions made during the channel model development in 3GPP RAN1 which may not be obvious and may not be documented in the official report [3].

So far, the evaluation and standardization of MIMO techniques in 3GPP has been primarily based on 2D channel models from SCM, ITU and WINNERII [4][5][6]. These models assume a 2D plane for the location of scatterers, reflectors and transmit/receive antennas, etc. This 2D assumption limits the MIMO transmission techniques (beamforming, precoding, spatial multiplexing, multi-user MIMO (MU-MIMO), etc.) to the azimuth dimension. In order to evaluate techniques such as user equipment (UE, aka mobile) specific elevation beamforming and full-dimension MIMO (FD-MIMO) [8], where the transmission is adapted efficiently in both elevation and azimuth to a particular UE, or vertical sectorization where a narrow elevation beam is tailored to each vertical sector, a 3D channel model is necessary. The 3D channel model described here is a geometry-based stochastic (GSCM) model (following the cluster-based approach common to COST-259, 273, 2100 as well as SCM, WINNERII models) and naturally extends the 2D channel models from ITU/WINNERII. It is also inspired by the extension from 2D to 3D channel model published as part of WINNERII/WINNER+ [6][7].

This article is organized as follows: In Section II we describe the target environments and the scope of the 3D channel model. Section I addresses the antenna model for a 2D polarized antenna array that can be used for simulations using the 3D channel model. The subsequent sections describe two key aspects of the 3D channel model – pathloss and LOS probability considering UE locations in high floors (Section IV) and modelling of a multi-path component in 3D (Section V). A summary of 3D channel model features from WINNERII, WINNER+ and 3GPP is provided in Section VI.

## II. APPLICABILITY OF THE 3D CHANNEL MODEL

The first step in the 3D channel modelling is to identify application environments. Urban Macro (3D-UMa) and Urban Micro (3D-UMi) with enhanced node B's (eNBs, aka base stations) located outdoors are considered to be typical usage scenarios for elevation beamforming and FD-MIMO. Using active antenna systems (AAS) in outdoor eNBs can provide better services and user experience for indoor as well as outdoor UEs. The usage of elevation beamforming and FD-MIMO for indoor eNBs (or distributed antenna systems – DAS) was not prioritized due to the cost of AAS and the reduced scope of adaptability in the elevation dimension in some indoor environments. The 3D-UMa and the 3D-UMi scenarios follow the conventional 2D-UMa and the 2D-UMi scenarios as determined in ITU-R [5]. Both scenarios are considered to be densely populated by buildings and homogeneous in nature – in terms of building height and building density. It may be assumed that the building blocks (in 3D-UMa and 3D-UMi scenarios) form a regular Manhattan type grid or can have a more irregular distribution while the building heights are typically distributed between 4 and 8 floors. In the case of 3D-UMa, it is assumed that the eNB height (25m) is well above the heights of the surrounding buildings so that over the rooftop diffractions form the dominant propagation mechanism for both indoor and outdoor UEs. In the case of 3D-UMi it is assumed that the eNB height (10m) is well below the heights of the surrounding buildings and therefore the received signal strength at the UE include contributions from both over the rooftop and around building propagation mechanisms. Note that the building density, building heights as well as the street layout described above were considered for both ray-tracing simulations and field measurements in order to derive the 3D channel models presented in this article.

The 3D channel model is applicable to UE heights ranging from 1.5m (street level) to 22.5m. A statistical approach for placing UEs is suggested for 3D-UMa and 3D-UMi that does not require modelling building dimensions explicitly. An indoor UE can be associated with a height given by $h_{UE} = 3(n_{fl} - 1) + 1.5$, where $n_{fl}$ is the floor number uniformly distributed between 1 and $N_{fl}$. $N_{fl}$ denotes the building height measured in floors and is uniformly distributed between 4 and 8. Outdoor UEs can be assumed to be at 1.5m height. This is a key aspect that extends the existing 3GPP [4] and ITU-R [5] methodologies where a UE is always modelled at the street level.

The 3D channel model is applicable to carrier frequencies between 2-6 GHz with up to 100 MHz of bandwidth. Higher carrier frequencies, e.g., up to 300 GHz, are of interest in 5G wireless communications and present a different set of challenges which are outside the scope of this 3D channel model. It also assumes that the size of the antenna array is negligible compared to the correlation distances of the large scale parameters such as shadow fading, delay spread, angle spread and Rician factor. These assumptions follow the practices of other existing channel models [5][6][7].

# III. ANTENNA MODELLING

Now that simulation environments are described some specific details of the 3D channel can be explained with a logical starting point being the eNB antennas themselves. A conventional deployment of a multi-antenna array at a macro eNB may use one or more cross-polarized antenna panels with +/-45 degree polarizations. Within each antenna panel multiple antenna elements per polarization are arranged in the vertical dimension, e.g., 8 elements, to concentrate the transmission within a narrow beamwidth in the vertical dimension (a half-power beamwidth on the order of 10 degrees). All of the antenna elements in the panel are used for transmission and reception but only one (logical) antenna port has been typically modelled along with a 2D channel model for system design and evaluation. As an example, a 4-port cross-polarized multiple-antenna array has been modelled as a set of 2 pairs of antenna ports arranged in the azimuth dimension, each pair comprising of a +45 and a -45 degree polarized co-located antenna ports. This arrangement provides an azimuth-only logical representation of a 2D antenna array comprising of 32 antenna elements (4 elements per row and 8 elements per column). Thus the study of MIMO techniques such as UE specific beamforming, MU-MIMO, etc. was limited to spatial adaptation in the azimuth dimension only.

The 3D channel model allows us to overcome this limitation and can be used for generating channel responses to each of the 32 antenna elements in the 2D array. Towards this end, a model for an individual antenna element is required for the 3D channel model.

*Polarized antenna element modelling*

In the interest of simplicity an antenna element at an eNB is characterized by an idealized parabolic antenna pattern with $65^0$ half-power beamwidth (in both azimuth and elevation) with 8 dBi antenna element gain. Polarization can be simulated by two models – a constant polarization model and a slanted dipole polarization model. A constant polarization model assumes that the polarization power split is independent of UE location (i.e., UE azimuth and elevation angles relative to broadside to the eNB array). A slanted dipole polarization model is based on the idea that a polarization slant can be modelled as a mechanical tilt. Considering a +45/-45 degree cross-polarized transmit antenna pair the constant polarization model assumption leads to an equal power split in vertical and horizontal directions for all UE locations. The slanted dipole polarization model achieves equal power split in vertical and horizontal directions at the antenna boresight but the power split ratio depends on the UE location in both azimuth and elevation dimensions. The choice of which polarization model to use in a particular simulation would be based on the expected antenna patterns of the eNBs being simulated. However it is worth noting that in detailed system simulations little difference is seen in the performance between the two polarization models.

# IV. LOS PROBABILITY AND PATHLOSS

With the basic antenna modelling done the next consideration before the fast fading modelling is LOS/NLOS state and pathloss determination. In the existing 3GPP [4] or ITU methodologies [5] the LOS probability model considers mainly street level UEs and the dependency on UE height is not explicitly considered. The impact of varying UE heights from 1.5m to 22.5m on LOS probability and pathloss modelling was studied for both 3D-UMa and 3D-UMi scenarios.

It may be recognized that one approach of modelling LOS probability, pathloss as well as fast-fading channel characteristics is to incorporate explicit building dimensions (a simplified ray-tracing approach) in the model. In 3GPP, however, it was decided to retain the fully stochastic approach of modelling as used in SCM [4] and in WINNER II [6] that does not depend on building/street dimensions explicitly. This enables reusing partly the existing modelling parameters from 2D stochastic models, helps incorporating measurement results from multiple sources while also reducing the complexity and processing time for system-level simulations. Also note that for UEs located indoors, the pathloss is simply determined as a sum of an outdoor pathloss component, wall penetration loss and an indoor pathloss component. Following this methodology, a LOS/NLOS state is associated to an indoor UE which means that the LOS/NLOS state is actually applicable to the outdoor pathloss component for that UE.

*LOS Probability*

The LOS probability for the 3D-UMa scenario is modelled as a sum of two probabilities - type-1 and type-2 LOS probabilities [11]. The decomposition of the overall LOS probability into type-1 and type-2 components is for modelling simplification. As shown in Figure 1(a) a UE is considered to be in type-1 LOS state if a UE in the first floor of the same building is also in a LOS state (UEs in the first floor can only be in type-1 LOS state). The type-1 LOS probability depends only on the horizontal distance between the eNB and the UE and follows the formula defined in the ITU model [5]. A LOS UE on a high floor of a building is considered to be in type-2 LOS state if a UE on the first floor of the same building can never achieve a LOS state. Note that the buildings are assumed to be at least 4-story high in 3D-UMa and hence type-2 LOS cases occur only when the UE is located on a floor higher than a 4-story building i.e., 12 m. Beyond 12m the probability to be in type-2 LOS state progressively increases with UE height and so is the overall LOS probability. The LOS probability model, which is the sum of the type-1 and type-2 LOS probabilities, for 3D-UMa as a function of UE height and distance is illustrated in Figure 3.

In a 3D-UMi scenario, where the eNB antenna is lower than the surrounding buildings type 1 and type 2 LOS states can be similarly defined as shown in Figure 1(b). However it was found by ray-tracing simulations that the UE height is not likely to affect LOS probability in the 3D-UMi scenario since type 2 LOS condition is limited to situations where the UE height is significantly higher than that of the blocking buildings and hence it can rarely occur, (mainly due to the low eNB height). Therefore the LOS probability model from ITU 2D-UMi [5] is reused for 3D-UMi scenario.

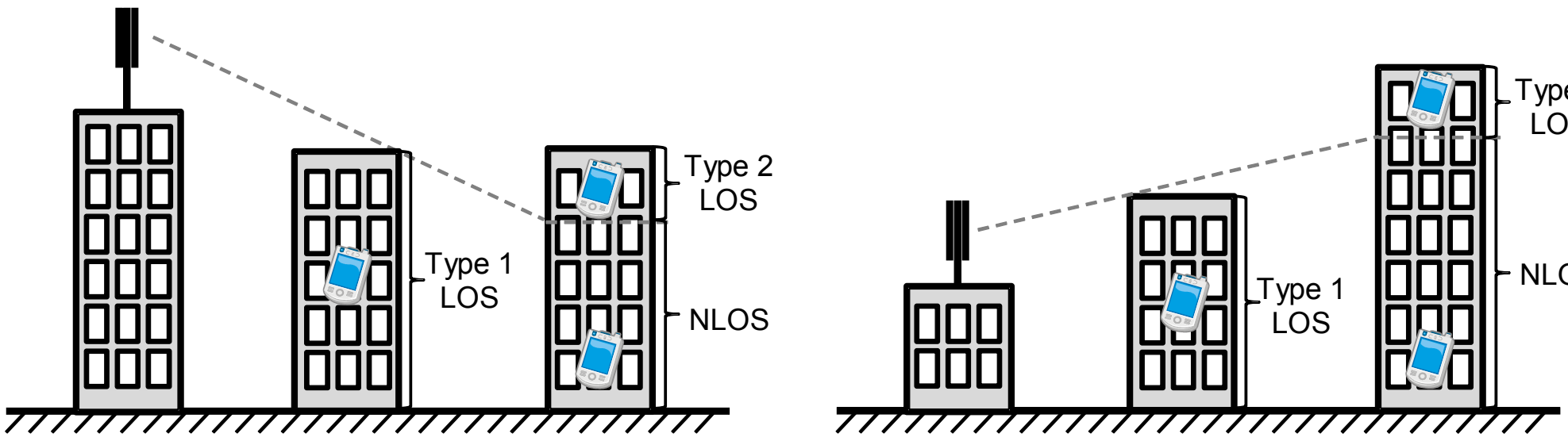

(a) 3D-UMa scenario (b) 3D-UMi scenario

**Figure 1: Type 1 and type 2 LOS probabilities**

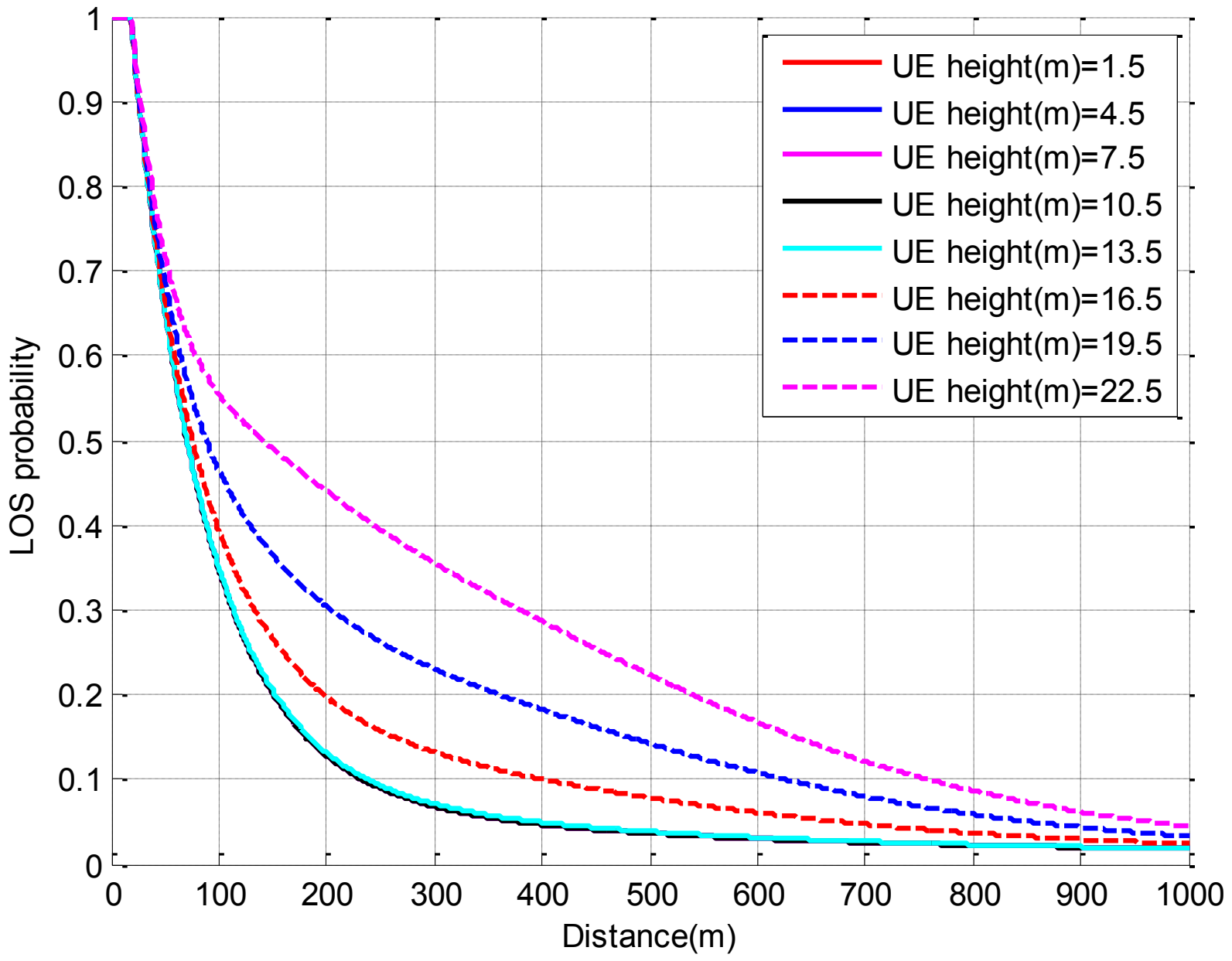

**Figure 2: Distance and height dependent LOS probability model for 3D-UMa**

*Pathloss Modelling (LOS)*

In [12] it is shown that a height-dependent pathloss for an indoor UE associated with a LOS state can be modelled considering the dimensions of the building and the location of the UE inside the building. In 3GPP it was agreed to model LOS pathloss by using the 3D distance between the eNB and the UE ($d_{3D}$) along with the coefficients given by the ITU LOS pathloss equations for both 3D-UMa and 3D-UMi. This provides a reasonable approximation to the more accurate model in [12] and can be determined without modelling building dimensions explicitly [13]. The ITU LOS pathloss model assumes a two-ray model resulting in a pathloss equation transitioning from a 22dB/decade slope to a steeper slope at a breakpoint depending on an environmental height. The environmental height represents the height of a dominant reflection from the ground (or a car) that can add constructively or destructively to the direct ray received at a UE located at the street level. In the 3D-UMa scenario, it is likely that such a dominant reflection path may come from the street-level for indoor UEs associated with a type-1 LOS condition. Therefore the environmental height is fixed at 1m for a UE associated with a type-1 LOS condition. In the case of a UE associated with a type-2 LOS condition, a dominant reflection can likely bounce of the rooftop of a neighbouring building. Noting that a rooftop is at least 12m in height, in this case the environmental height is randomly determined from a discrete uniform distribution of (12m, 15m,…, ($h$-1.5)m) where $h$ is the UE height in meter.

*Pathloss Modelling (NLOS)*

For NLOS pathloss modelling, Figure 3(a) shows the primary radio propagation mechanism in a 3D-UMa environment where the dominant propagation paths travel via multiple-diffraction over rooftops followed by diffraction at the edge of a building. The pathloss attenuation increases with the diffraction angle as a UE transitions from a high floor to a low floor. In order to model this phenomenon a linear height gain term given by $-\alpha(h-1.5)$ is introduced, where $\alpha$ (in dB/m) is the gain coefficient. A range of values between 0.6 and 1.5 were observed in different results based on field measurements and ray tracing simulations and eventually a value of 0.6 dB/m was agreed (see section 7.2.7.1 in [16] for references). Figure 3(b) shows the principle of radio propagation for 3D-UMi environment where the dominant propagation paths travel through and around buildings. The UE may also receive small energy from propagation above rooftops. Considering simplicity of modelling, a linear height gain is also applied to the 3D-UMi NLOS pathloss with a 0.3 dB/m gain coefficient based on the results from multiple sources. In addition, for both 3D-UMa and 3D-UMi scenarios, the NLOS pathloss is lower-bounded by the corresponding LOS pathloss because the pathloss in a NLOS environment is in principle larger than that in a LOS environment.

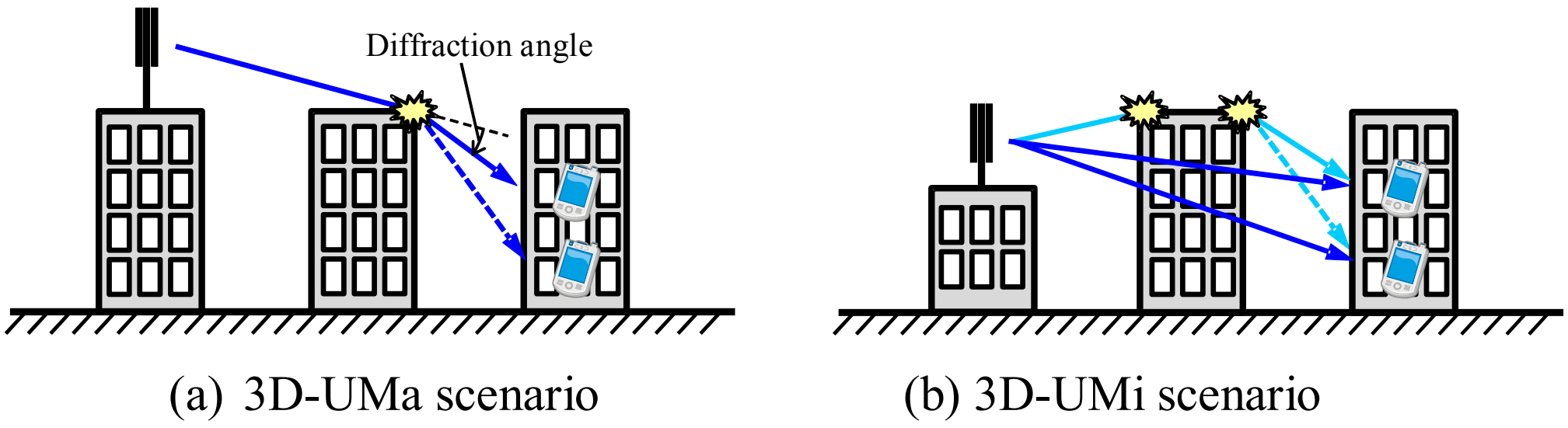

**Figure 3: Principle of the radio propagation for NLOS environment**

# V. FAST FADING MODEL

Now that antenna modelling, LOS probability and pathloss have been described fast fading model can be readily presented in this section. In the following, a cellular downlink is assumed for describing the fast fading model and hence the departure angles are defined at the eNB side and the arrival angles are defined at the UE side. The fast fading channel coefficients model the time-varying fluctuations of wireless channels that are caused by the combination of multipath and UE movement. The channel coefficients of a link between a transmitter and a receiver are determined by the composite channel impulse responses of the multiple path components (MPCs). Each MPC is characterized by a path delay, a path power and random phases introduced during the propagation as well as the incident path angles, i.e., azimuth angles of departure and arrival (AOD and AOA) and zenith angles of departure and arrival (ZOD and ZOA).

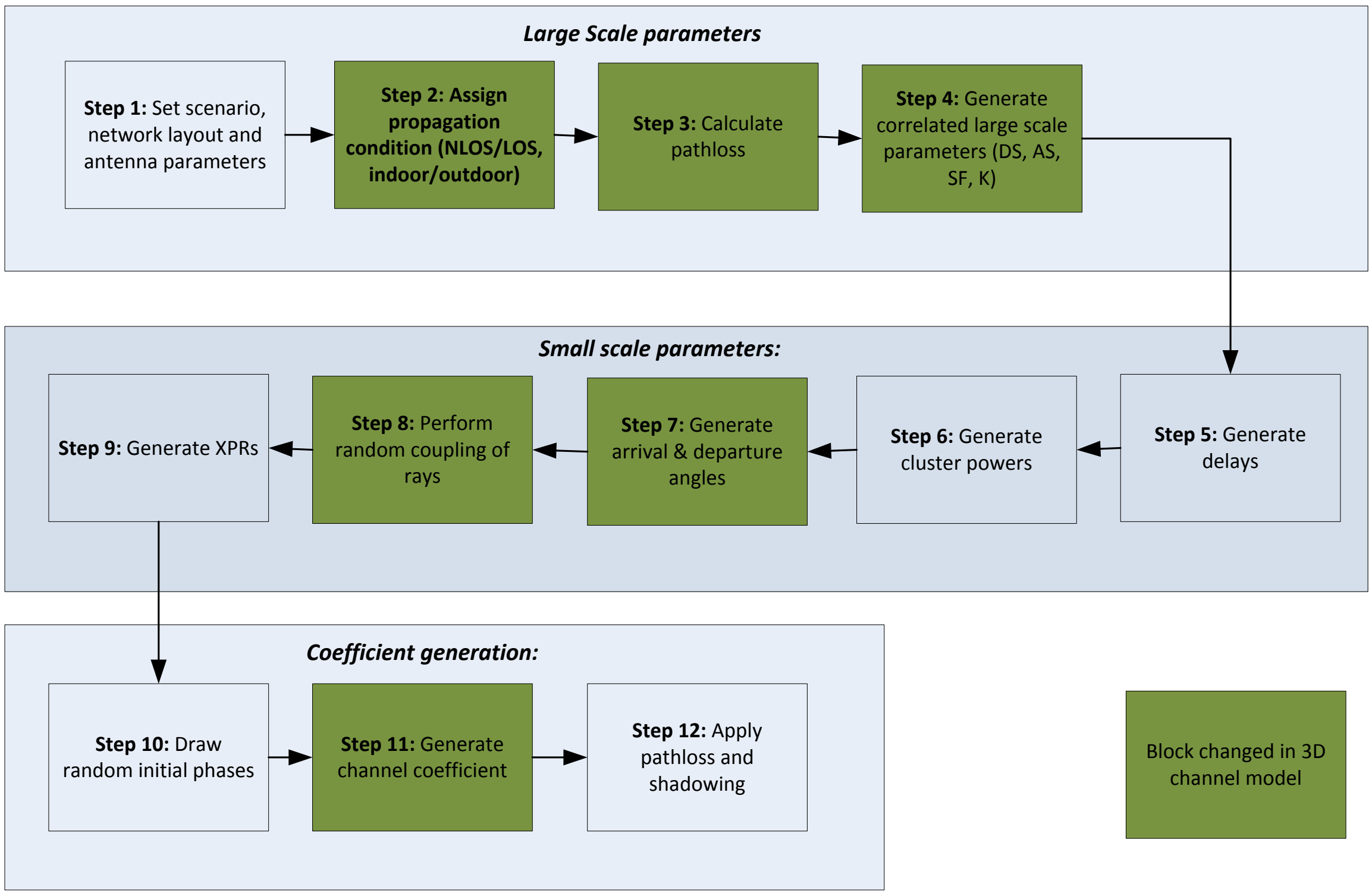

**(a) Channel generation steps in the 3GPP 3D channel model [3]**

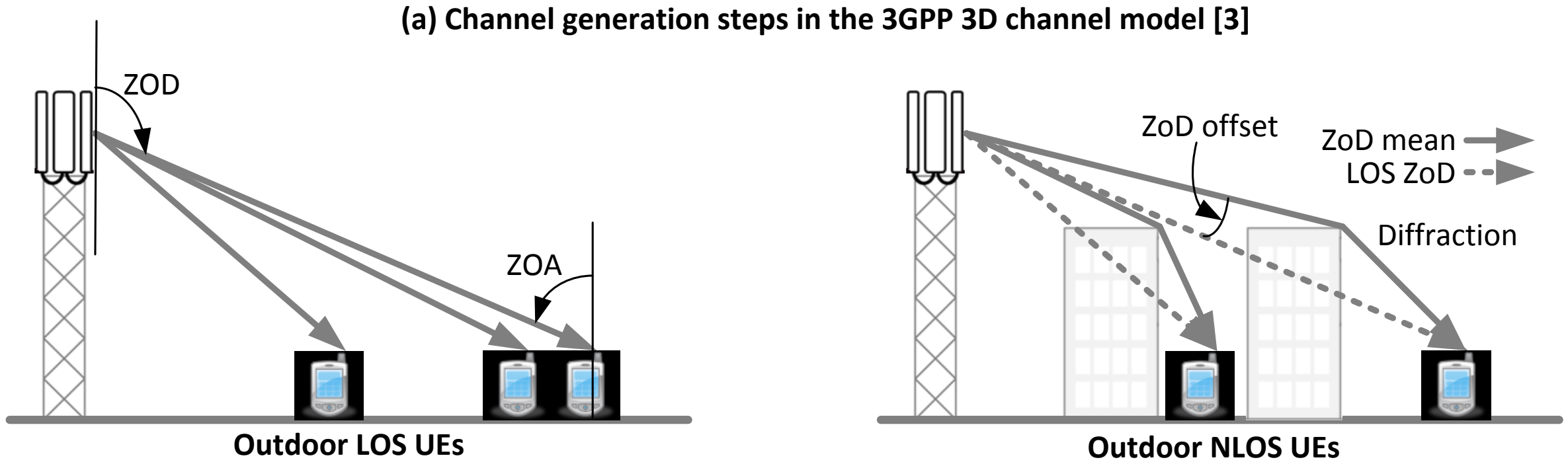

**(b) ZOD in outdoor LOS and NLOS conditions**

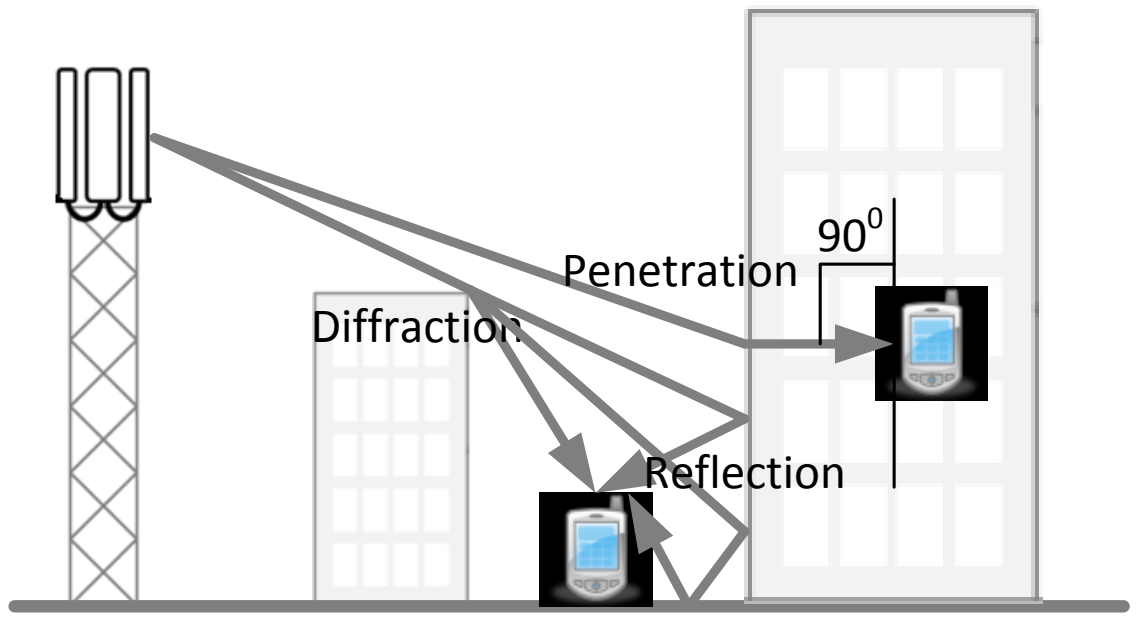

**(c) ZOA in indoor and outdoor NLOS conditions**

**Figure 4: Fast-fading channel generation steps and propagation mechanisms**

Figure 4(a) shows the steps to generate fast fading channel coefficients in the 3GPP 3D channel model. The high-level procedure shown in Figure 4(a) is the same as its precedent 2D channel model in TR 36.814 [9] but a few steps have been revised to take into account the channel characteristics in the elevation domain. As explained in Section IV a LOS or NLOS state associated to a UE is jointly determined in step 2 by both UE horizontal distance and UE height. The pathloss in step 3 is determined based upon a LOS probability and a height-gain term depending on the UE height and distance from the eNB. In step 4 and step 7 zenith angle spreads at departure and at arrival (ZSD and ZSA) and ZOD and ZOA are generated in addition to the azimuth spread values and angles ASD, ASA, AOD and AOA. In step 8, for each path, the AOA sub-paths are randomly coupled with the AOD sub-paths, the ZOA sub-paths and the ZOD sub-paths. Finally, in step 11, the channel generation equation is modified to take into account ZOAs and ZODs. The rest of this section explains in more detail the newly-introduced elevation parameters and procedures in these revised steps in the 3GPP 3D channel models.

*Power angular spectrum in zenith (PAS-Z)*

Measurement results and ray tracing data have indicated that the marginal distribution of the composite power angular spectrum in zenith (PAS-Z) is Laplacian, and its conditional distribution given a certain link distance and UE height can also be approximated by Laplacian [14][15]. To incorporate these observations ZOD and ZOA are modeled by inverse Laplacian functions.

*Composite ZSD*

It is also observed that the ZSD decreases significantly as a UE moves further away from the eNB [14]. An intuitive explanation is provided by the fact that the angle subtended by a fixed local ring of scatterers at the UE to the eNB decreases as the UE moves away from the eNB. The ZSD is also observed to change to a smaller extent as an indoor UE moves up to higher floors [14][15]. The ZSD model incorporated in 3GPP is a function of UE height and distance from the eNB as shown in Figure 5.1 and Figure 5.2. Applying these ZSD models to 3D-UMa with 500m inter site distance (ISD) and 3D-UMi with 200m ISD described in Section II with UE height distributed between 1.5 and 22.5m we can observe the overall distribution of ZSD in a wireless network deployment. In Figure 5.5 we show the empirical probability density of ZSD for the serving cell links as observed in the 3D-UMa and 3D-UMi scenarios. Note that the ZSD tends to be a little smaller in the case of 3D-UMa reflecting a clear dominance of above roof-top propagation mechanism.

*ZOD offset*

Note that in the azimuth dimension the AODs (corresponding to different MPCs) for a given link are centered at the azimuth LOS angle between the UE and the eNB. In the elevation dimension, however, the ZODs are not centered at the LOS zenith angle for NLOS cases. This can be observed from Figure 4(b) NLOS case where it is shown that the mean ZOD is shifted from the LOS ZOD because the bulk of the energy is received via diffraction from the building edges (rooftops) blocking the direct path between the UE and the eNB. The magnitude of the ZOD offset becomes smaller as the distance of the link increases partly because the angle subtended by a given building to an eNB decreases with distance. It is also clear from Figure 4(b) NLOS case that the ZOD offset decreases with an increase in UE height. These observations were captured in the ZOD offset model shown in Figure 5.3 and 5.4. In the case of 3D-UMi the dependence of ZOD offset on UE height was not modeled because the results showed some variation across different sources (see Table 7.3-8 in [3] for references). On the other hand, it may be noted that when an indoor UE is associated with a LOS state (denoted as LOS outdoor to indoor, or LOS O-to-I), the effect of diffraction paths does not dominate and hence the ZOD offset is modeled to be zero. The application of the ZOD offset models from Figure 5.3 and 5.4 to the 3D-UMa and 3D-UMi scenarios with UE height ranging from 1.5 to 22.5m results in mean ZOD distributions as shown in Figure 5.6. It may be noted that for 3D-UMi the eNB height is below the surrounding buildings leading to mean ZOD distributed on both sides of the horizon (90 degrees) while for 3D-UMa the eNB height is above the surrounding buildings leading to mean ZOD distributed on only one side of the horizon.

The ZOAs (corresponding to different MPCs) for UEs located indoors is modeled to be centered at 90 degrees. This is justified especially when the UE is deep inside the building where the indoor paths are guided by the floor and the ceiling as Figure 4 (c) illustrates. The ZOA center angle for outdoor UEs is modeled as the LOS ZOD angle for simplicity.

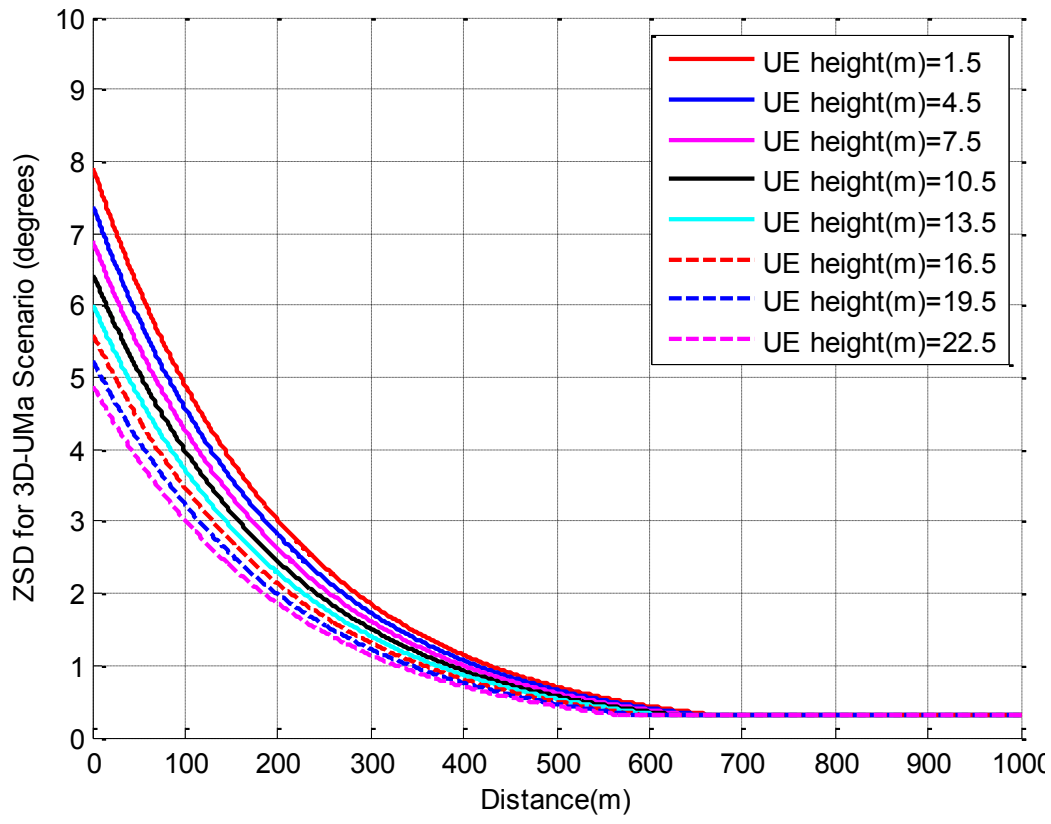

**Figure 5.1: ZSD model for 3D-UMa**

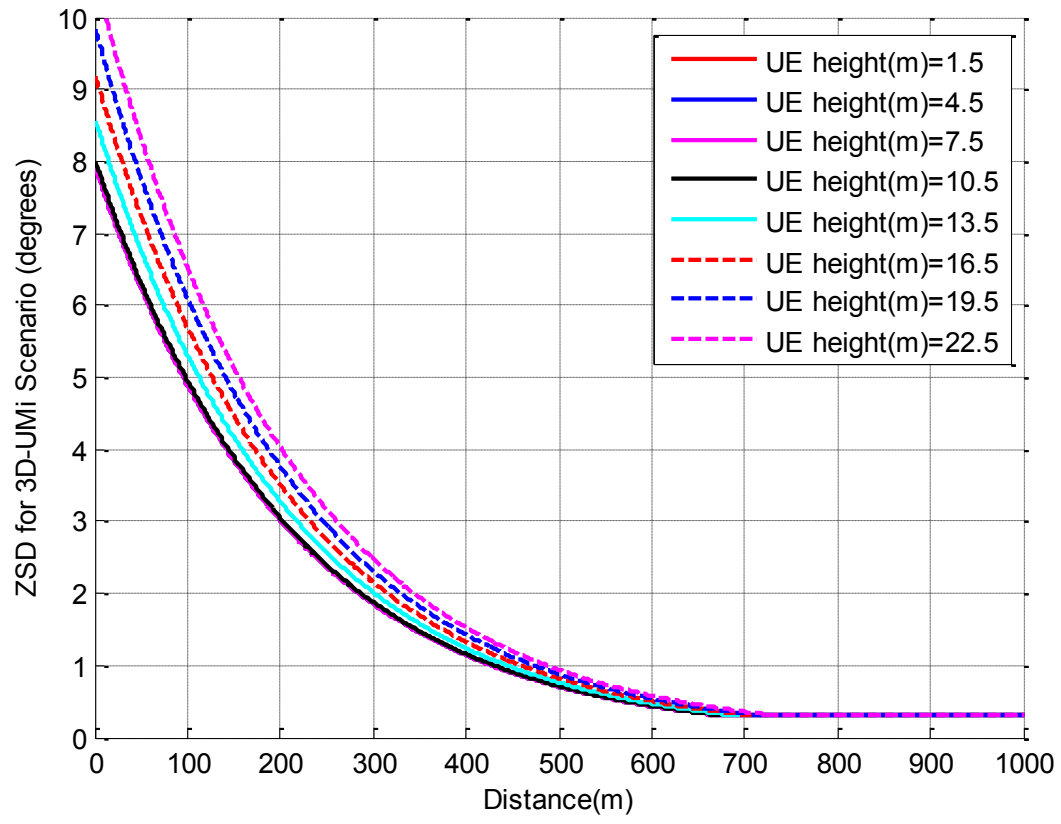

**Figure 5.2:ZSD model for 3D-UMi**

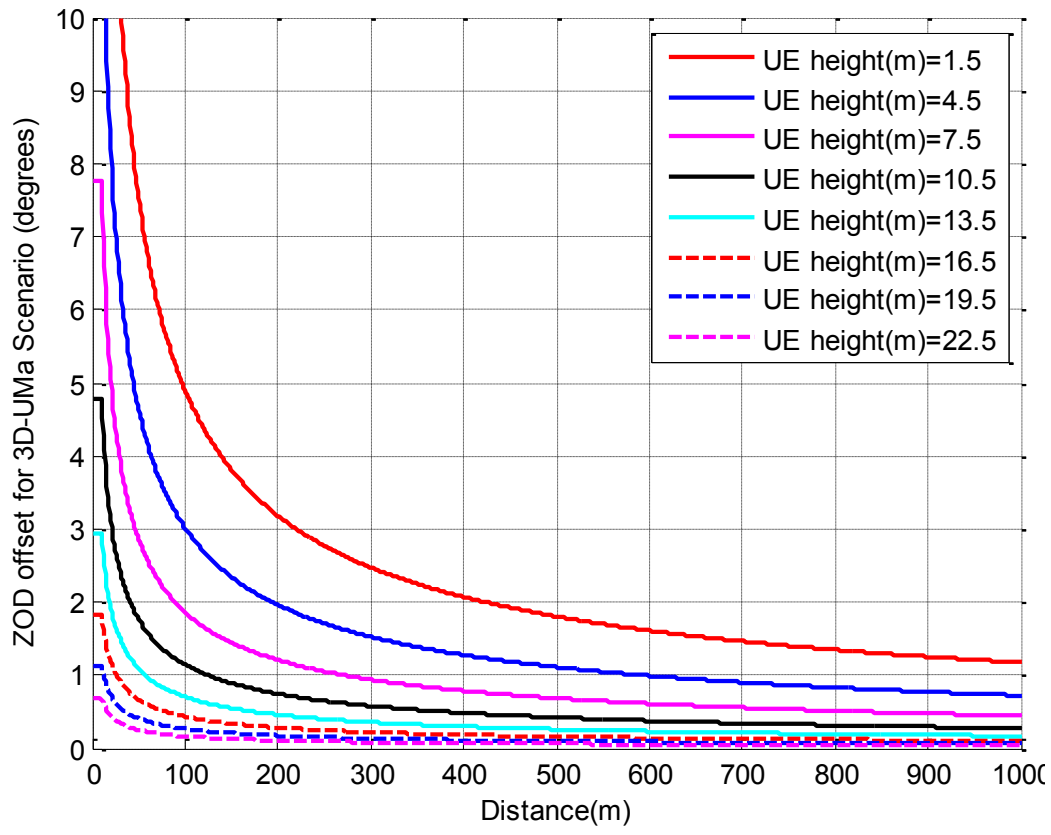

**Figure 5.3: Model for the magnitude of ZOD offset for 3D-UMa**

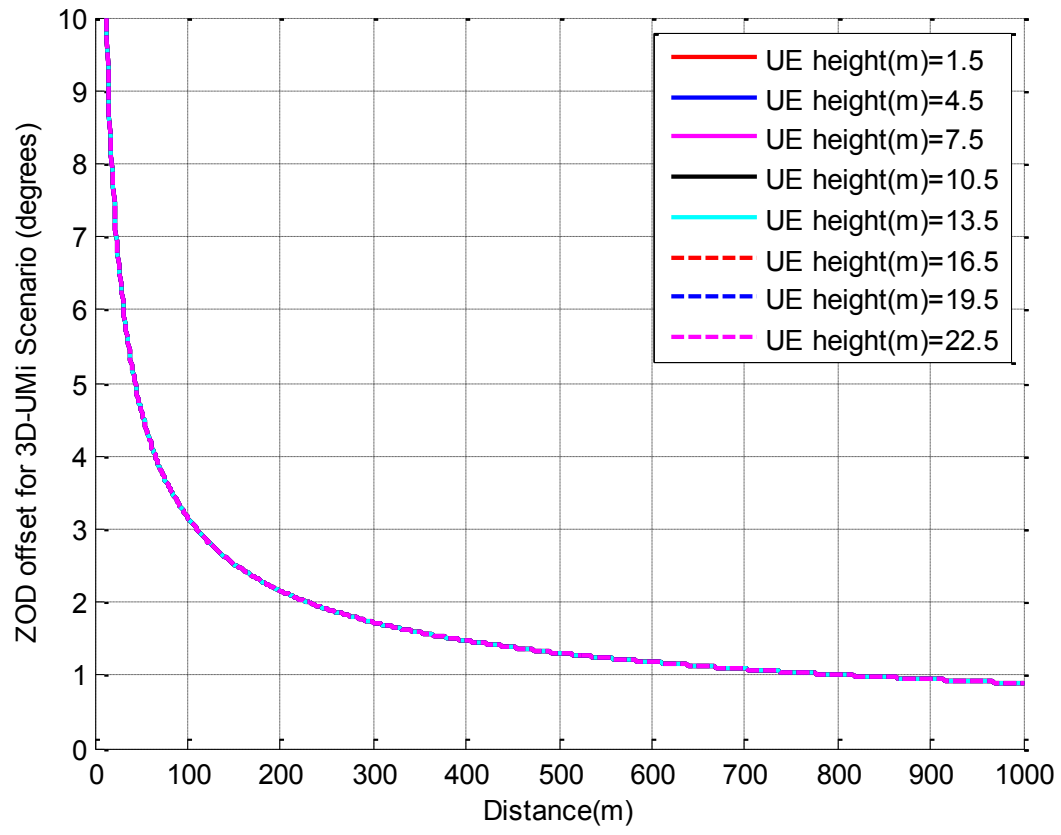

**Figure 5.4: Model for the magnitude of ZOD offset for 3D-UMi**

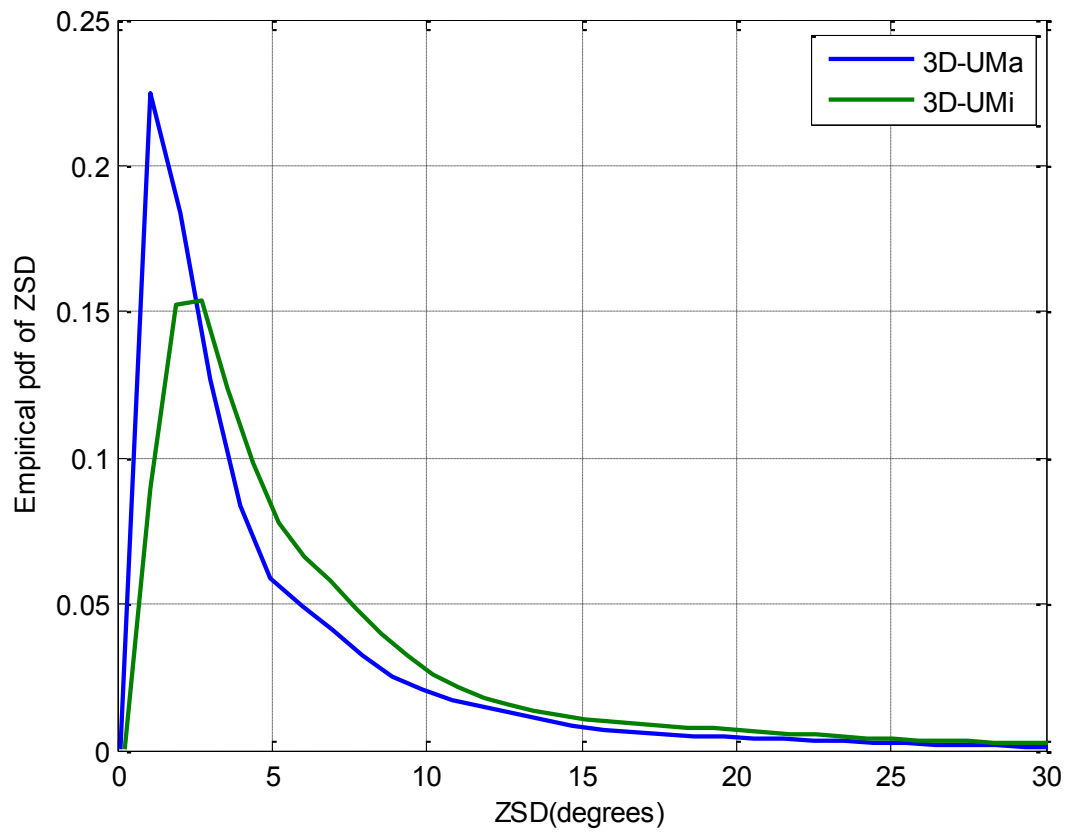

**Figure 5.5: Empirical pdf of ZSD (for the serving cell links) as observed in 3D-UMa and 3D-UMi scenarios**

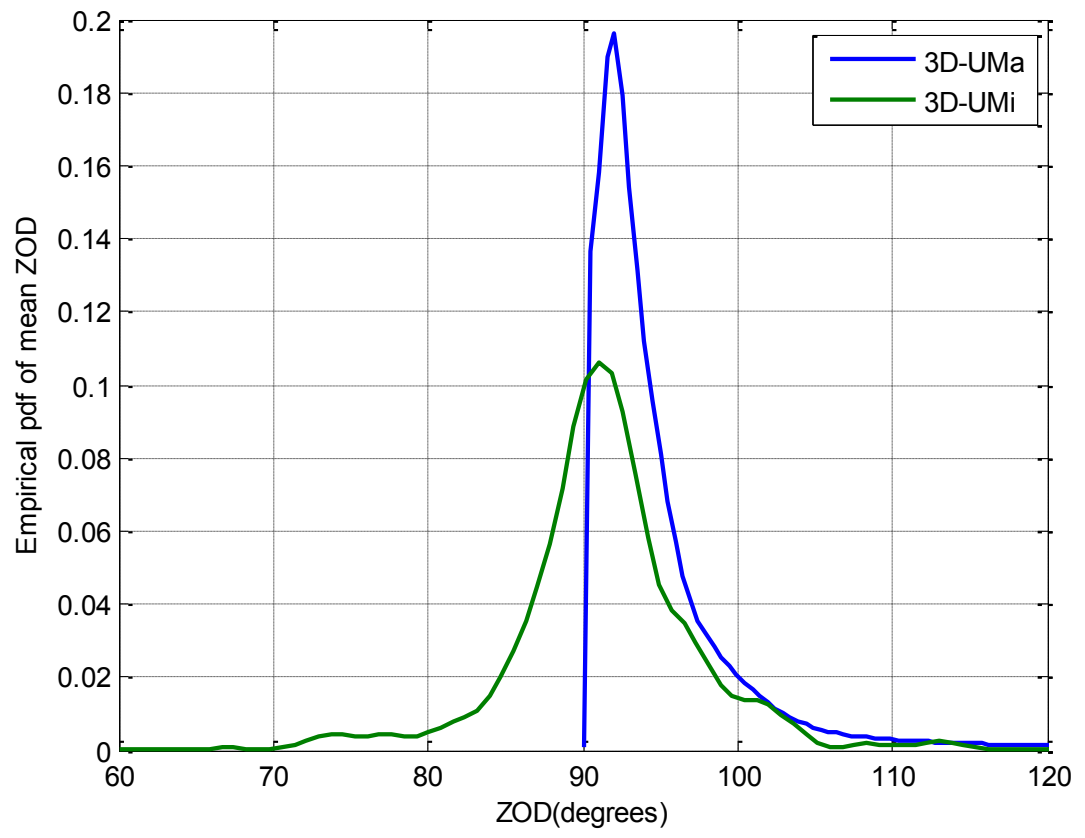

**Figure 5.6: Empirical pdf of mean ZOD (for the serving cell links) as observed in 3D-UMa and 3D-UMi scenarios ($90^0$ is horizon)**

*Fast fading channel generation equation*

After characterizing all azimuth and zenith angles of MPCs and related parameters, the channel of the $n$-th path between $u$-th UE antenna and $s$-th eNB antenna is given by:

$$H_{u,s,n}(t) = \sqrt{P_n/M}\sum_{m=1}^{M}\begin{bmatrix} F_{rx,u,\theta}(\theta_{n,m,ZOA},\phi_{n,m,AOA}) \\ F_{rx,u,\phi}(\theta_{n,m,ZOA},\phi_{n,m,AOA}) \end{bmatrix}^T \begin{bmatrix} \exp\left(j\Phi_{n,m}^{\theta\theta}\right) & \sqrt{\kappa_{n,m}^{-1}}\exp\left(j\Phi_{n,m}^{\theta\phi}\right) \\ \sqrt{\kappa_{n,m}^{-1}}\exp\left(j\Phi_{n,m}^{\phi\theta}\right) & \exp\left(j\Phi_{n,m}^{\phi\phi}\right) \end{bmatrix}$$

$$\begin{bmatrix} F_{tx,s,\theta}(\theta_{n,m,ZOD},\phi_{n,m,AOD}) \\ F_{tx,s,\phi}(\theta_{n,m,ZOD},\phi_{n,m,AOD}) \end{bmatrix} \exp\left(j2\pi\lambda_0^{-1}\left(\hat{r}_{rx,n,m}^T.\bar{d}_{rx,u}\right)\right)\exp\left(j2\pi\lambda_0^{-1}\left(\hat{r}_{tx,n,m}^T.\bar{d}_{tx,s}\right)\right)\exp\left(j2\pi v_{n,m}t\right)$$

Here, $F_{rx,u,\theta}$ and $F_{rx,u,\phi}$ are field patterns of the receive antenna element $u$ in the direction of the spherical basis vectors, a zenith basis vector $\hat{\theta}$ and an azimuth basis vector $\hat{\phi}$, respectively. $F_{tx,s,\theta}$ and $F_{tx,s,\phi}$ are field patterns of the transmit antenna element $s$ in the direction of $\hat{\theta}$ and $\hat{\phi}$, respectively. The 2x2 matrix between the arrival and departure field pattern vectors is the depolarization matrix, of which two diagonal elements characterise the co-polarized phase response, and the two off-diagonal elements model the cross-polarized phase response and power attenuation $\sqrt{\kappa_{n,m}^{-1}}$. If polarisation is not considered, the 2x2 polarisation matrix can be replaced by the scalar $\exp\left(j\Phi_{n,m}\right)$ and only vertically polarised field patterns are applied. $\hat{r}_{rx,n,m}$ is the spherical unit vector with azimuth arrival angle $\phi_{n,m,AOA}$ and zenith arrival angle $\theta_{n,m,ZOA}$, given by:

$$\hat{r}_{rx,n,m} = \begin{bmatrix} \sin\theta_{n,m,ZOA}\cos\phi_{n,m,AOA} \\ \sin\theta_{n,m,ZOA}\sin\phi_{n,m,AOA} \\ \cos\theta_{n,m,ZOA} \end{bmatrix};$$

and $\hat{r}_{tx,n,m}$ is the spherical unit vector with azimuth departure angle $\phi_{n,m,AOD}$ and elevation departure angle $\theta_{n,m,ZOD}$, given by

$$\hat{r}_{tx,n,m} = \begin{bmatrix} \sin\theta_{n,m,ZOD}\cos\phi_{n,m,AOD} \\ \sin\theta_{n,m,ZOD}\sin\phi_{n,m,AOD} \\ \cos\theta_{n,m,ZOD} \end{bmatrix}.$$

$\bar{d}_{rx,u}$ and $\bar{d}_{tx,s}$ are respectively the location vectors of receive antenna element $u$ and transmit antenna element s. $\lambda_0$ is the wavelength of the carrier frequency.

# VI. SUMMARY/CONCLUSIONS

A 3D channel model is an integral part of the evaluation of multi-antenna techniques exploiting elevation domain channel characteristics for supporting higher spectral efficiency. In this article, we provided a summary and some insight into the 3D channel model development in 3GPP spanning approximately a year starting in January 2013. We described the modelling of UEs at high-floors that is necessary to understand elevation beamforming and FD-MIMO performance. The height-gain model used to extend the 2D pathloss models to 3D was motivated and the height dependence of LOS probability was described. The methodology of extending the 2D fast-fading model from WINNERII/ITU to 3D was described and the dependence of the new elevation parameters on UE height and distance from the eNB was illustrated. A comparison of the 3D channel model features from WINNERII, WINNER+ and 3GPP is provided in Table 1.

**Table 1: High-level comparison of 3D channel models from WINNERII, WINNER+ and 3GPP.**

| | WINNERII[6] | WINNER+[7] | 3GPP TR36.873[3] |
|---|---|---|---|
| Scenarios | Indoor, indoor to outdoor, outdoor to indoor (macro, micro), urban | Indoor, outdoor, outdoor to indoor (macro, micro), urban, suburban | Outdoor, outdoor to indoor (macro, micro), urban |
| eNB height (outdoor) | 25m (macro), 10m (micro) | 25m (macro), 10m (micro) | 25m (macro), 10m (micro) |
| UE height (indoor) | 1.5 – 7.5m (3 floors) | 1.5 – 7.5m (3 floors) | 1.5 - 22.5m (8 floors) |
| LOS probability | Not height dependent | Not height dependent | Height dependent |
| Pathloss | No height-gain | Height-gain modelled | Height-gain modelled |
| Power angular spectrum (zenith) | Gaussian | Laplacian | Laplacian |
| Mean ZSD | Constant | Constant | Distance and height dependent |
| ZOD offset | Not modelled | Constant | Distance and height dependent |

# VII. ACKNOWLEDGEMENT

The authors from Nokia would like to acknowledge the help from Huan Nguyen from Nokia, Aalborg and Mark Schamberger from Nokia, USA.

Note – 3GPP documents may be downloaded from ftp://ftp.3gpp.org. WINNER II, WINNER+ documents may be downloaded from http://projects.celtic-initiative.org/winner+/index.html.

## ABBREVIATIONS

The following abbreviations are used in this article:

3D-UMa: 3-Dimensional Urban-Macro scenario
3D-UMi: 3-Dimensional Urban-Micro scenario
3GPP: Third Generation Partnership Project
AAS: Advanced Antenna System
AOA: Azimuth Angle of Arrival
AOD: Azimuth Angle of Departure
ASA: Azimuth Angular Spread at Arrival
ASD: Azimuth Angular Spread at Departure
COST: European Cooperation in Science and Technology
DAS: Distributed Antenna System
eNB: enhanced Node B aka base-station
FD-MIMO: Full Dimension Multiple Input Multiple Output
GSCM: Geometry based Stochastic Channel Model
IMT: International Mobile Telecommunications

ISD: Inter-Site Distance
ITU: International Telecommunication Union
ITU-R: International Telecommunication Union - Radio communication sector
LOS: Line-of-Sight
MIMO: Multiple Input Multiple Output
MPC: Multi-Path Component
MU-MIMO: Multi-User MIMO
NLOS: Non Line-of-Sight
PAS-Z: Power Angular Spectrum in Zenith
RAN1: Radio Access Network Working Group for Radio Layer 1
SCM: 3GPP Spatial Channel Model
UE: User Equipment
WINNER: Wireless Word Initiative New Radio
ZOA: Zenith Angle of Arrival
ZOD: Zenith Angle of Departure
ZSA: Zenith Angular Spread at Arrival
ZSD: Zenith Angular Spread at Departure

## BIOGRAPHIES

Bishwarup Mondal received the B.E. and M.E. degrees from Jadavpur University and the Indian Institute of Science in 1997 and 2000 respectively and the Ph.D. degree from the University of Texas at Austin in 2006. He is presently with Nokia, Arlington Heights, IL. He is the recipient of the 2005 IEEE Vehicular Technology Society Daniel E. Noble Fellowship and a co-author of the best student paper award in IEEE Globecom 2006.

Eugene Visotsky received a B.S., an M.S. and a Ph.D. in Electrical Engineering in 1996, 1998, and 2000, respectively, from University of Illinois at Urbana-Champaign. He is currently with Nokia, Arlington Heights, IL. His current areas of interest are in advanced inter-cell interference coordination, cooperative transmission algorithms and 3D MIMO techniques.

Frederick W. Vook (SM'04) received the B.S. degree from Syracuse University in 1987 and the M.S. and Ph.D. degrees from The Ohio State University in 1989 and 1992, respectively, all in electrical engineering. From 1992 to 2011, he was with Motorola, Schaumburg, IL. Since 2011, he has been with Nokia, Arlington Heights, IL, where his current work involves advanced antenna array solutions for LTE and 5G cellular systems.

Timothy A. Thomas received the B.S. degree from the University of Illinois at Urbana-Champaign in 1989; the M.S. degree from the University of Michigan, in 1990; and the Ph.D. degree from Purdue University in 1997. From 1997 to 2011, he was with Motorola, Schaumburg, IL, and since 2011, he has been with the North American Radio Systems Research Group, Technology and Innovation Office, Nokia, Arlington Heights, IL.

Amitabha (Amitava) Ghosh joined Motorola in 1990 after receiving his Ph.D in Electrical Engineering from Southern Methodist University, Dallas. Since joining Motorola he worked on multiple wireless technologies starting from IS-95, cdma-2000, 1xEV-DV/1XTREME, 1xEV-DO, UMTS, HSPA, 802.16e/WiMAX/802.16m, Enhanced EDGE and 3GPP LTE. Dr. Ghosh has 60 issued patents and numerous external and internal technical papers. Currently, he is Head, North America Radio Systems Research within the Technology and Innovation office of Nokia Networks. He is currently working on 3GPP LTE-Advanced and 5G technologies. His research interests are in the area of digital communications, signal processing and wireless communications. He is a senior member of IEEE and co-author of the book titled "Essentials of LTE and LTE-A".

YOUNG-HAN NAM received his B.S. and M.S from Seoul National University, Korea, in 1998 and 2002, respectively. He received a Ph.D. in electrical engineering from the Ohio State University, Columbus, in 2008. Since February 2008 he has been working at Samsung Research America at Dallas, Richardson, Texas. He has been engaged in standardization, design and analysis of the 3GPP LTE and LTE-Advanced from Release 8 through 12. His research interests include MIMO/multi-user/cooperative wireless communications, cross-layer design, and information theory.

Yang Li is a Senior Research Engineer in Samsung Research America, Dallas. He received a B.S. and a M.S. degree, in 2005 and 2008, in electronic engineering from Shanghai Jiao Tong University, Shanghai, China, and a Ph.D. degree in 2012, in electrical engineering from The University of Texas at Dallas, Richardson, TX, USA. His research interests include MIMO, interference management, cooperative communication and cognitive radio.

Jianzhong (Charlie) Zhang [S'96, M'02, SM'09]: Jianzhong(Charlie) Zhang is currently senior director and head of Wireless Communications Lab with Samsung Research America at Dallas, where he leads technology development, prototyping and standardization for Beyond 4G and 5G wireless systems. From Aug 2009 to Aug 2013, he served as the Vice Chairman of the 3GPP RAN1 working group and led development of LTE and LTE-Advanced technologies such as 3D channel modeling, UL-MIMO and CoMP, Carrier Aggregation for TD-LTE, etc. Before joining Samsung, he was with Motorola from 2006 to 2007 working on 3GPP HSPA standards, and with Nokia Research Center from 2001 to 2006 working on IEEE 802.16e (WiMAX) standard and EDGE/CDMA receiver algorithms. He received his Ph.D. degree from University of Wisconsin, Madison.

Min Zhang received his M.S. degree with Distinction in Telecommunication from University of Canterbury, New Zealand in 2005, and Ph.D. degree from Australian National University in Telecommunication in 2009. He worked for Alcatel-Lucent since 2009 as wireless researcher and senior standards engineer, where he is currently working on 3GPP LTE/LTE-A standardisation. His interests include wireless channel modelling, communication and information theory, MIMO-OFDM system design, and performance optimization of standards.

Qinglin Luo is a leading research scientist with Bell Labs Shanghai. His primary responsibilities include conducting and leading research projects on air-interface technologies for next generation communications. In previous postings, he has been with Motorola, UK as a base station system architect, and with China Academy of Telecommunication Technologies (CATT), as a standard engineer. He holds a Ph.D. degree in electrical engineering from University of Surrey, UK. His current research interests include advanced antenna array techniques, massive MIMO, error control coding, wireless cloud, etc.

Yuichi Kakishima received the B.S. and M.S. degrees from Tokyo Institute of Technology, Tokyo, Japan, in 2005 and 2007, respectively. In 2007, he joined NTT DOCOMO, INC. Since joining NTT DOCOMO, he has been engaged in the research and development of wireless access technologies including multiple-antenna transmission techniques for LTE and LTE-Advanced systems.

Koshiro Kitao was born in Tottori, Japan, in 1971. He received B.S., M.S. and Ph. D. degrees from Tottori University, Tottori, Japan in 1994, 1996 and 2009 respectively. He joined the Wireless Systems Laboratories, Nippon Telegraph and Telephone Corporation (NTT), Kanagawa, Japan, in 1996. Since then, he has been engaged in the research of radio propagation for mobile communications. He is now Research Engineer of the Research Laboratories, NTT DOCOMO, INC., Kanagawa, Japan.